\documentclass[amsmath,jcp,preprint]{revtex4-1}
\usepackage{color}
\usepackage{amsfonts}
\usepackage{amssymb}
\usepackage{graphicx}

\begin{document}
\title{On the widths of Stokes lines in Raman scattering from molecules adsorbed at metal surfaces and in molecular conduction junctions}
\author{Yi Gao}
\email{yig057@ucsd.edu}
\affiliation{Department of Chemistry \& Biochemistry, University of California San Diego, La Jolla CA 92093, USA}
\author{Michael Galperin}
\email{migalperin@ucsd.edu}
\affiliation{Department of Chemistry \& Biochemistry, University of California San Diego, La Jolla CA 92093, USA}
\author{Abraham Nitzan}
\email{nitzan@post.tau.ac.il}
\affiliation{Department of Chemistry, University of Pennsylvania, Philadelphia, PA 19104, USA}
\affiliation{School of Chemistry, Tel Aviv University, Tel Aviv, 69978, Israel}

\begin{abstract}
Within a generic model we analyze the Stokes linewidth 
in surface enhanced Raman scattering (SERS) from molecules embedded as bridges
in molecular junctions. We identify four main contributions to the off-resonant 
Stokes signal and show that under zero voltage bias 
(a situation pertaining also to standard SERS experiments)
and at low bias junctions only one of these contributions is pronounced.
The  linewidth of this component is determined by the molecular vibrational
relaxation rate, which is dominated by interactions with the essentially bosonic 
thermal environment when the relevant molecular electronic energy is far from 
the metal(s) Fermi energy(ies). It increases when the molecular electronic level is
close to the metal Fermi level so that an additional vibrational relaxation channel due to electron-hole (eh) excition in the molecule opens. Other contributions to the Raman signal, 
of considerably broader linewidths, can become important at larger junction bias.
\end{abstract}

\maketitle

\section{Introduction}\label{intro}
Molecular optoelectronics is an active field of research made possible 
by advances in laser technology and nanofabrication.\cite{MGANPCCP12}
The possibility to conduct optical measurements in open non-equilibrium nano-systems
resulted in the appearance of new diagnostic tools,
and offers a route to optical control schemes such as 
switching in molecular electronics devices.
Standard observables of optical spectroscopy can yield new information when
monitored in open current-carrying molecular junctions. 
For example, current-induced fluorescence\cite{HoPRB08,HoPRL10}
yields information on molecular resonances in the non-equilibrium system  
and makes imaging at submolecular resolution feasible, while the  
intensity of the emitted light corresponds to charge current noise
at optical frequencies\cite{BerndtPRL10,BerndtPRL12}
and can yield information on fast voltage transients at the tunnel junction.\cite{LothAPL13}
Raman spectroscopy of current-carrying junctions 
can serve as a diagnostic tool similar to inelastic tunneling spectroscopy,
and as an indicator for current-induced heating of electronic and vibrational degrees of freedom.\cite{CheshnovskySelzerNatNano08,NatelsonNL08,NatelsonNatNano11}
(Possible pitfalls of such characterization were discussed 
theoretically~\cite{MGANJPCL11,MGANPRB11}).
Recently, measurements of {\em dc} current and/or noise in response
to laser pulse pair sequence was suggested as a variant of pump-probe 
spectroscopy for molecular junctions capable of providing information
on intra-molecular dynamics at sup-picosecond timescale.\cite{SelzerPeskinJPCC13,OchoaSelzerPeskinMGJPCL15}

As noted above, the ability to characterize vibrational structure of a molecular device
makes Raman scattering similar to inelastic electron tunneling spectroscopy (IETS). 
The corresponding spectra are characterized by their peak positions and heights, 
as well as lineshapes and linewidths. 
In addition to standard peaks, rich IETS lineshape features caused by 
interference between elastic and inelastic scattering channels are known.\cite{MGRatnerNitzanJCP04,RaiMGPRB12}  
Similar interference features in Raman scattering were recently discussed.\cite{ApkarianMGANPRB16}
The dependence of (resonant) IETS spectra on gate and source-drain biases
was measured and discussed.\cite{McEuenNature00,McEuenNL05,YaoNL06} 
It appears to primarily manifest the sensitivity of molecular
normal modes to the molecule charging state.\cite{GalperinNitzanRatnerPRB08,WhiteGalperinPCCP12}
Similarly, a shift in the frequencies of Stokes lines with bias was 
observed\cite{NatelsonNatNano11,NatelsonPCCP13}  
and was shown to result at least partly from the voltage dependence of 
the charge on the molecule.\cite{KaasbjergNitzanPRB13,KronikNeatonNatelsonPNAS14,WhiteTretiakNL14,WhiteOchoaMGJPCC14}
Finally, the linewidths of IETS signals where studied both 
experimentally\cite{ReedNL04} and theoretically\cite{MGRatnerANNL04}
and were shown to be dominated by the strength of electron-phonon interactions.
No such study has been done so far for Raman scattering from
molecular junctions.

The present paper focuses on the latter issue:
we identify the main contribution to the observed Raman intensity and analyze, 
using a generic model, the non-monotonic dependence of the Stokes linewidth on 
the gate and bias potentials.
In Section~\ref{model} we introduce our model for an illuminated molecular junction
as well as our calculation methodology for off-resonant
Raman scattering from this system. 
Section~\ref{numres} presents our results and Section~\ref{conclude} concludes.

%%%%%%%%%%%%%%%%%%%%%%%%%%%%%%%%%%%%%%%%%%%%%%%%%%%%%%%%%%%%%%%%%%%%%%%%%%%%%%%
\begin{figure}[b]
\centering\includegraphics[width=\linewidth]{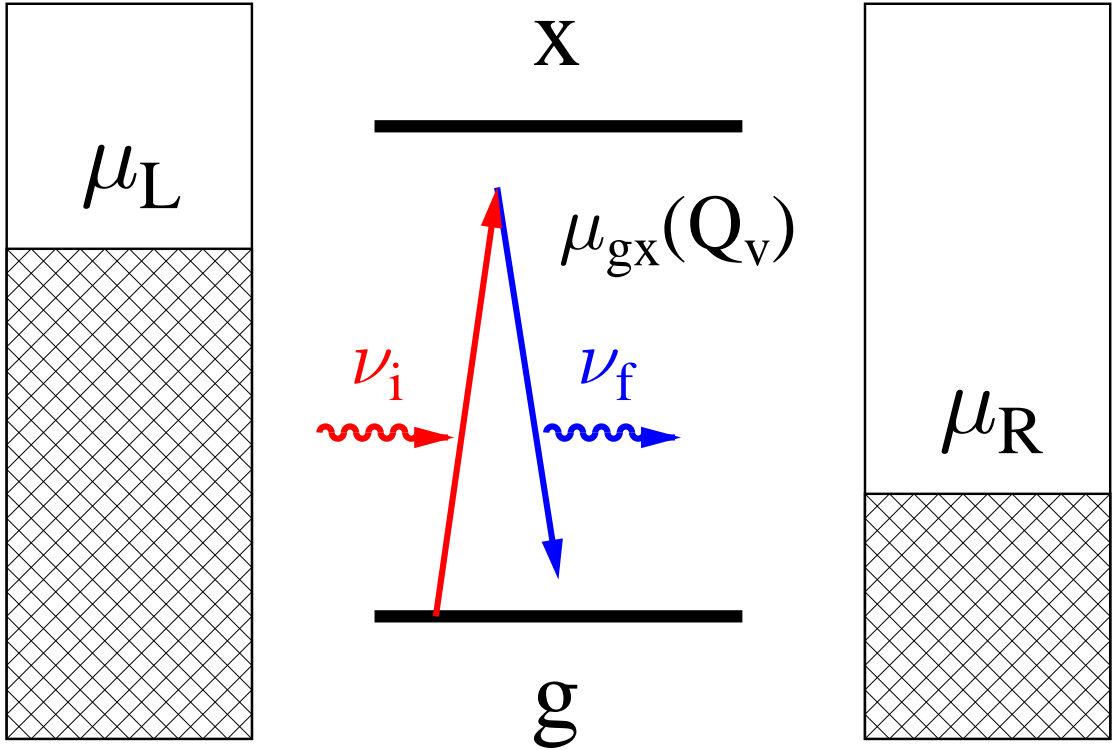}
\caption{\label{fig1}
(Color online) A sketch of the model for off-resonant Raman scattering
in a junction.
}
\end{figure}
%%%%%%%%%%%%%%%%%%%%%%%%%%%%%%%%%%%%%%%%%%%%%%%%%%%%%%%%%%%%%%%%%%%%%%%%%%%%%%%

\section{Model and Method}\label{model}
We consider junction comprised of a molecule coupled to two metallic contacts,
$L$ and $R$, each at its own equilibrium. 
The molecule is represented by two electronic levels
(ground, $\varepsilon_g$, and excited, $\varepsilon_x$, states) 
and a molecular vibration, taken harmonic of frequency $\omega_v$, 
linearly coupled to the levels populations (an Holstein-type model). 
The junction is subjected to an external radiation field,
represented by a set of quantum harmonic modes $\{\nu_\alpha\}$
 (see sketch in Fig.~\ref{fig1}). 
One of these modes, of frequency $\nu_i$ represent the incident mode that pumps 
the system. All other modes, $\{\nu_f\}$, are taken to be vacant. 
The Hamiltonian is  
\begin{equation}
\label{H}
\hat H = \hat H_0 + \hat H_{rad} + \hat V 
\end{equation}
where $\hat H_0$ represents the dark junction,
$\hat H_{rad}$ is Hamiltonian of the radiation field, 
and $\hat V$ is the molecule-field coupling. Explicitly
\begin{align}
\label{H0}
&\hat H_0 = \sum_{m=g,x}\varepsilon_m\hat n_m
+ \omega_v \hat v^\dagger\hat v + \sum_k\varepsilon_k\hat n_k
+ \sum_\beta \omega_\beta \hat b^\dagger_\beta\hat b_\beta
\nonumber \\
&\qquad + \sum_{k,m}\bigg(V_{km}\hat c_k^\dagger\hat d_m+ H.c.\bigg) 
+\sum_{m=g,x} M_m\hat Q_v \hat n_m
\\
&\qquad +\sum_{beta} V^{th}_\beta \hat Q_\beta\hat Q_v
\nonumber \\
&\hat H_{rad} =  \sum_{\alpha\in i\{f\}} \nu_\alpha\hat a_\alpha^\dagger\hat a_\alpha
\\
\label{V}
&\hat V = \sum_{\alpha}\bigg(U_{\alpha D}(\hat Q_v)\,\hat a_\alpha^\dagger\hat D
+H.c.\bigg)
\end{align}
Here $\hat d_m^\dagger$ ($\hat d_m$) and $\hat c_k^\dagger$ ($\hat c_k$)
create (annihilate) electrons in the molecular level $m$ and state $k$ of the 
metal contacts, respectively. 
$\hat n_m=\hat d_m^\dagger\hat d_m$ and $\hat n_k=\hat c_k^\dagger\hat c_k$
are the corresponding electron number operators for states 
$m$ (${}=g,x$) of the molecule and $k$ of the contacts. 
$\hat D^\dagger=\hat d_x^\dagger\hat d_g$ and $\hat D=\hat d_g^\dagger\hat d_x$ 
are molecular excitation and de-excitation operators.
$\hat v^\dagger$ ($\hat v$) and $\hat b_\beta^\dagger$ ($\hat b_\beta$) 
create (annihilate) vibrational quanta in the molecule and mode $\beta$ of the thermal bath, 
respectively. $\hat Q_v=\hat v+\hat v^\dagger$ and 
$\hat Q_v=\hat b_\beta+\hat b_\beta^\dagger$ are the oscillators position operators.
$\hat a_\alpha^\dagger$ ($\hat a_\alpha$) 
creates (destroys) photon in the mode $\alpha$ of radiation field. 
Note that this model contains two interactions that can cause inelastic light
scattering. First is the dependence of the molecule-field coupling $U$ 
on the vibrational coordinate. The other is the polaronic coupling term in Eq.(\ref{H0}) 
whose importance is measured by the electron-vibration coupling $M$.

Following Refs.~\cite{GalperinRatnerNitzanNL09,GalperinRatnerNitzanJCP09} 
and focusing on the low voltage bias regime,
we consider only `normal Raman' scattering, i.e. a process where the initial state 
is its ground state.\footnote{For strongly biased junctions both lower and upper 
molecular states may be partially occupied, giving rise to more contributions to elastic 
and inelastic light scattering, see Ref.~\cite{GalperinRatnerNitzanJCP09}.} 

Raman scattering is a coherent process of fourth order in the matter-radiation field 
coupling (two orders correspond to the outgoing photon, blue line in Fig.~\ref{fig1}, 
and two orders correspond to the incoming photon, red line in Fig.~\ref{fig1}). 
Explicit steady-state expression for the `normal Raman'
scattering from the initial mode $i$ to a final mode $f$ of the radiation field is
(see Ref.~\cite{GalperinRatnerNitzanJCP09} for details)
\begin{align}
\label{Jif}
&J_{i\to f} = \int_{-\infty}^{+\infty}d(t'-t) \int_{-\infty}^0 d(t_1-t) \int_{-\infty}^0 d(t_2-t')\,
\nonumber \\ &
 e^{-i\nu_f(t'-t)}\, e^{-i\nu_i(t_1-t_2)}\times
\\ &
\left\langle \hat U_{iD}(t_2)\hat D(t_2)\, \hat U_{Df}(t')\hat D^\dagger(t')\,
\hat U_{f D}(t)\hat D(t)\,\hat U_{Di}(t_1)\hat D^\dagger(t_1)\right\rangle
\nonumber
\end{align}
where $\hat U_{\alpha D}\equiv U_{\alpha D}(\hat Q_v)$.
As in standard treatments, we expand the molecule-field coupling to linear term 
in Taylor series in the molecular vibrational displacement
\begin{equation}
\label{UTaylor}
U_{\alpha D}(\hat Q_v)\approx  U_{\alpha D}^{(0)}+U_{\alpha D}^{(1)}\,\hat Q_v
\end{equation}
Depending on combination of molecule-field coupling terms 
($U_{\alpha D}^{(0)}$ or $U_{\alpha D}^{(1)}\,\hat Q_v$)
in the expression (\ref{Jif}) one gets contributions to vibrational and electronic Raman 
(Rayleigh) scatterings. 
For example, substituting only $U_{\alpha D}^{(0)}$ in place of all
molecule-field couplings in Eq.(\ref{Jif}) yields the pure electronic Raman 
contribution discussed in Refs.~\cite{MGANJPCL11,MGANPRB11}. 
Here we focus on the vibrational Raman scattering, whose lowest order contribution 
comes from terms that are second order in the coupling to the molecular
vibration. Such terms will be of order $\left(U^{(1)}\right)^2$.
After collecting all such contributions to the vibrational Raman we 
(a)~separate vibrational and electronic degrees of freedom 
(i.e. neglecting vibration-induced electronic correlations)
and (b) neglect electronic correlation between ground
and excited states of the molecule assuming that the energy gap between them
is much larger than the widths associated with their coupling to the contacts. 
We focus on off-resonant Raman scattering and restrict our consideration to gate 
voltages that keep the upper electronic level above the leads chemical
potentials (so it is essentially unpopulated), 
%we also disregard the electron-phonon coupling at the level
%(taking $M_x=0$). 
Under these approximations the explicit expression becomes
\begin{widetext}
\begin{subequations}
\label{Jnu}
\begin{align}
%%%%%%%
\label{Jnuifa}
& J_{\nu_i\to\nu_f} = \rho(\nu_i)\Delta\nu_i\,\rho(\nu_f)\Delta\nu_f\mbox{Re}
\int\frac{dE_{g1}}{2\pi} \int\frac{dE_{g2}}{2\pi}
\int\frac{dE_{x1}}{2\pi} \int\frac{dE_{x2}}{2\pi}
\bigg\{
\nonumber \\ &
iD^{>}(\nu_{if}) G_g^{<}(E_{g1})G_g^{<}(E_{g2})G_x^{>}(E_{x1})G_x^{>}(E_{x2})
\frac{2\, U_{iD}^{(0)}U_{Df}^{(1)}U_{fD}^{(0)}U_{Di}^{(1)}
+U_{iD}^{(1)}U_{Df}^{(0)}U_{fD}^{(0)}U_{Di}^{(1)}
+U_{iD}^{(0)}U_{Df}^{(1)}U_{fD}^{(1)}U_{Di}^{(0)}}
{(\nu_f-E_{x2}+E_{g2}+i\delta)(\nu_i-E_{x1}+E_{g1}-i\delta)}
\end{align}
\begin{align}
%%%%%%%
\label{Jnuifb}
& -iD^{>}(\nu_{if}-E_{g21}) G_g^{<}(E_{g1})G_g^{>}(E_{g2})G_x^{>}(E_{x1})G_x^{>}(E_{x2})
\bigg( 
\frac{2\, U_{iD}^{(0)}U_{Df}^{(1)}U_{fD}^{(0)}U_{Di}^{(1)}}{(\nu_f-E_{x2}+E_{g2}+i\delta)(\nu_i-E_{x1}+E_{g1}-i\delta)}
 \\ & \qquad\quad +
\frac{U_{iD}^{(1)}U_{Df}^{(0)}U_{fD}^{(0)}U_{Di}^{(1)}}
{(\nu_f-E_{x2}+E_{g2}+i\delta)(\nu_f-E_{x1}+E_{g2}-i\delta)}
+
\frac{U_{iD}^{(0)}U_{Df}^{(1)}U_{fD}^{(1)}U_{Di}^{(0)}}
{(\nu_i-E_{x2}+E_{g1}+i\delta)(\nu_i-E_{x1}+E_{g1}-i\delta)}
\bigg)
\nonumber 
\end{align}
\begin{align}
%%%%%%%
\label{Jnuifc}
& -iD^{>}(\nu_{if}-E_{x21}) G_g^{<}(E_{g1})G_g^{<}(E_{g2})G_x^{<}(E_{x1})G_x^{>}(E_{x2})
\bigg( 
\frac{2\, U_{iD}^{(0)}U_{Df}^{(1)}U_{fD}^{(0)}U_{Di}^{(1)}}{(\nu_f-E_{x1}+E_{g2}+i\delta)(\nu_i-E_{x2}+E_{g1}-i\delta)}
 \\ & \qquad\quad +
\frac{U_{iD}^{(1)}U_{Df}^{(0)}U_{fD}^{(0)}U_{Di}^{(1)}}{(\nu_f-E_{x1}+E_{g2}+i\delta)(\nu_f-E_{x1}+E_{g1}-i\delta)}  +
\frac{U_{iD}^{(0)}U_{Df}^{(1)}U_{fD}^{(1)}U_{Di}^{(0)}}{(\nu_i-E_{x2}+E_{g2}+i\delta)(\nu_i-E_{x2}+E_{g1}-i\delta)} \bigg)
\nonumber
\end{align}
\begin{align}
\label{Jnuifd}
& -iD^{>}(\nu_{if}-E_{x21}-E_{g21}) 
G_g^{<}(E_{g1})G_g^{>}(E_{g2})G_x^{<}(E_{x1})G_x^{>}(E_{x2})
\bigg( 
\frac{2\, U_{iD}^{(0)}U_{Df}^{(1)}U_{fD}^{(0)}U_{Di}^{(1)}}{(\nu_f-E_{x1}+E_{g2}+i\delta)(\nu_i-E_{x2}+E_{g1}-i\delta)}
 \\ & \qquad\qquad\qquad\qquad\qquad\qquad\qquad\qquad +
\frac{U_{iD}^{(1)}U_{Df}^{(0)}U_{fD}^{(0)}U_{Di}^{(1)}}
{\lvert\nu_f-E_{x1}+E_{g2}+i\delta\rvert^2} +
\frac{U_{iD}^{(0)}U_{Df}^{(1)}U_{fD}^{(1)}U_{Di}^{(0)}}
{\lvert\nu_i-E_{x2}+E_{g1}+i\delta)\rvert^2}
\bigg)
\bigg\}
\nonumber 
\end{align}
\end{subequations}
\end{widetext}
Here $\nu_{if}=\nu_i-\nu_f$, $E_{m21}=E_{m2}-E_{m1}$ ($m=g,x$), 
$G_m^{>/</r}(E)$ and $D^{>}(\omega)$ are Fourier transforms of the greater/lesser/retarded projections of the single electron Green function and the greater projection of 
the phonon Green function, respectively
\begin{align}
 G_m(\tau,\tau') =& -i\langle T_c\,\hat d_m(\tau)\,\hat d_m^\dagger(\tau')\rangle
 \\
 D(\tau,\tau') =& -i\langle T_c\,\hat Q_v(\tau)\,\hat Q_v(\tau')\rangle
\end{align}
where $T_c$ is the contour ordering operator. 
$\rho(\nu)\equiv \nu^2/\pi^2c^3$ is the %radiative-coupling weighted %????
density of optical modes.  

%%%%%%%%%%%%%%%%%%%%%%%%%%%%%%%%%%%%%%%%%%%%%%%%%%%%%%%%%%%%%%%%%%%%%%%%%%%%%%%
\begin{figure}
\centering\includegraphics[width=\linewidth]{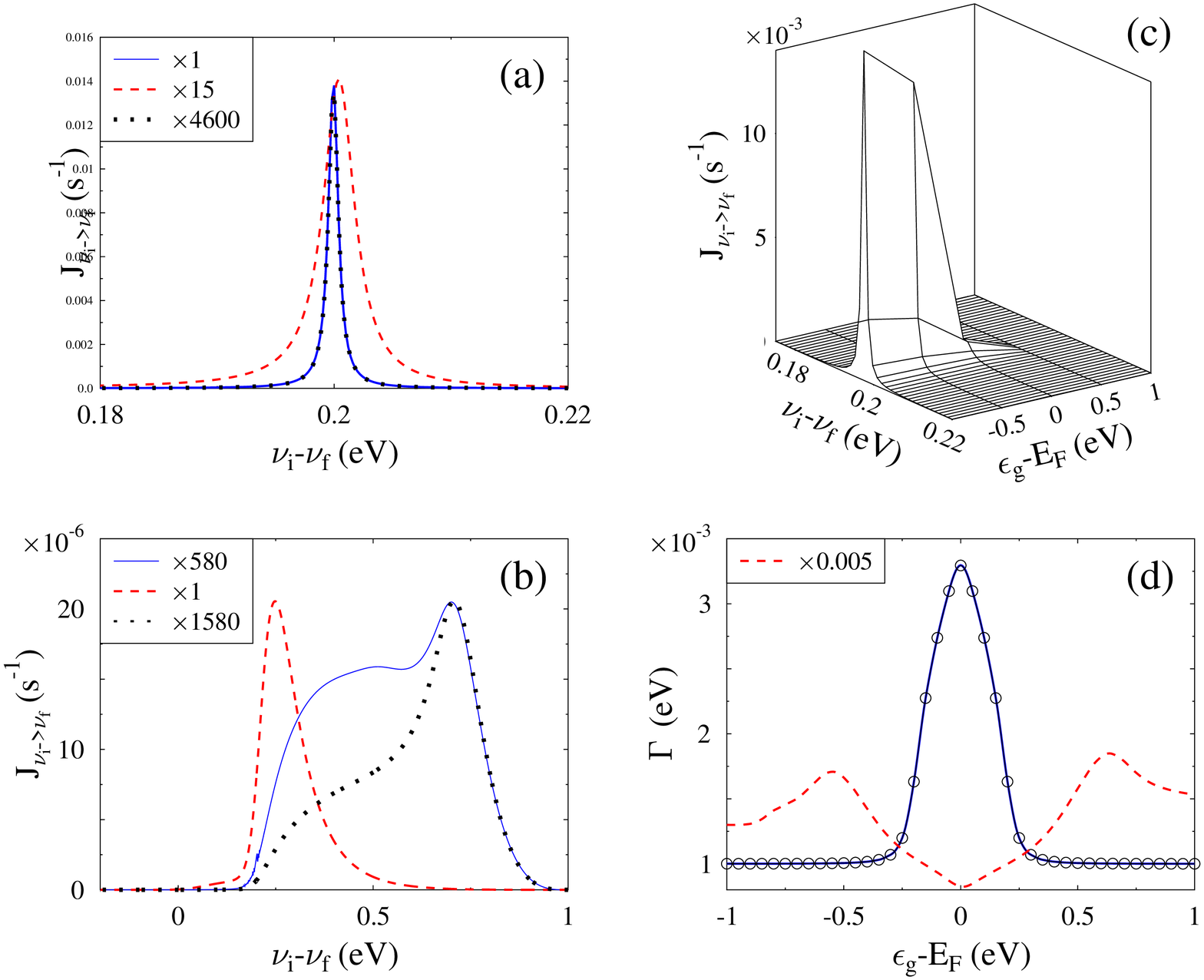}
\caption{\label{fig2}
(Color online) The Stokes component of the vibrational Raman scattering at equilibrium, 
$\mu_L=\mu_R=E_F$, for $\Gamma_m^K=0.05$~eV.
Shown are the contributions of (a)  Eq.~(\ref{Jnuifa}) and (b) Eq.~(\ref{Jnuifb}) vs. 
Raman shift  for three level positions: ($\varepsilon_g-E_F=-0.5$~eV, solid line, blue), 
at ($\varepsilon_g-E_F=0$, dashed line, red)
and above ($\varepsilon_g-E_F=0.5$~eV, dotted line, black).
The total Stokes signal, Eq.~(\ref{Jif}), as function of the Raman shift 
and level position is shown in panel (c). Panel (d) shows 
widths $\Delta\nu$ (standard deviations) of the two main contributions 
(Eq.~(\ref{Jnuifa}) - solid line, blue; Eq.~(\ref{Jnuifb}) - dashed line, red)  
as functions of the level position.
Circles indicate broadening of the molecular vibration due to coupling to electron-hole
excitations. 
See text for other parameters.
}
\end{figure}
%%%%%%%%%%%%%%%%%%%%%%%%%%%%%%%%%%%%%%%%%%%%%%%%%%%%%%%%%%%%%%%%%%%%%%%%%%%%%%%
%%%%%%%%%%%%%%%%%%%%%%%%%%%%%%%%%%%%%%%%%%%%%%%%%%%%%%%%%%%%%%%%%%%%%%%%%%%%%%%
\begin{figure}
\centering\includegraphics[width=\linewidth]{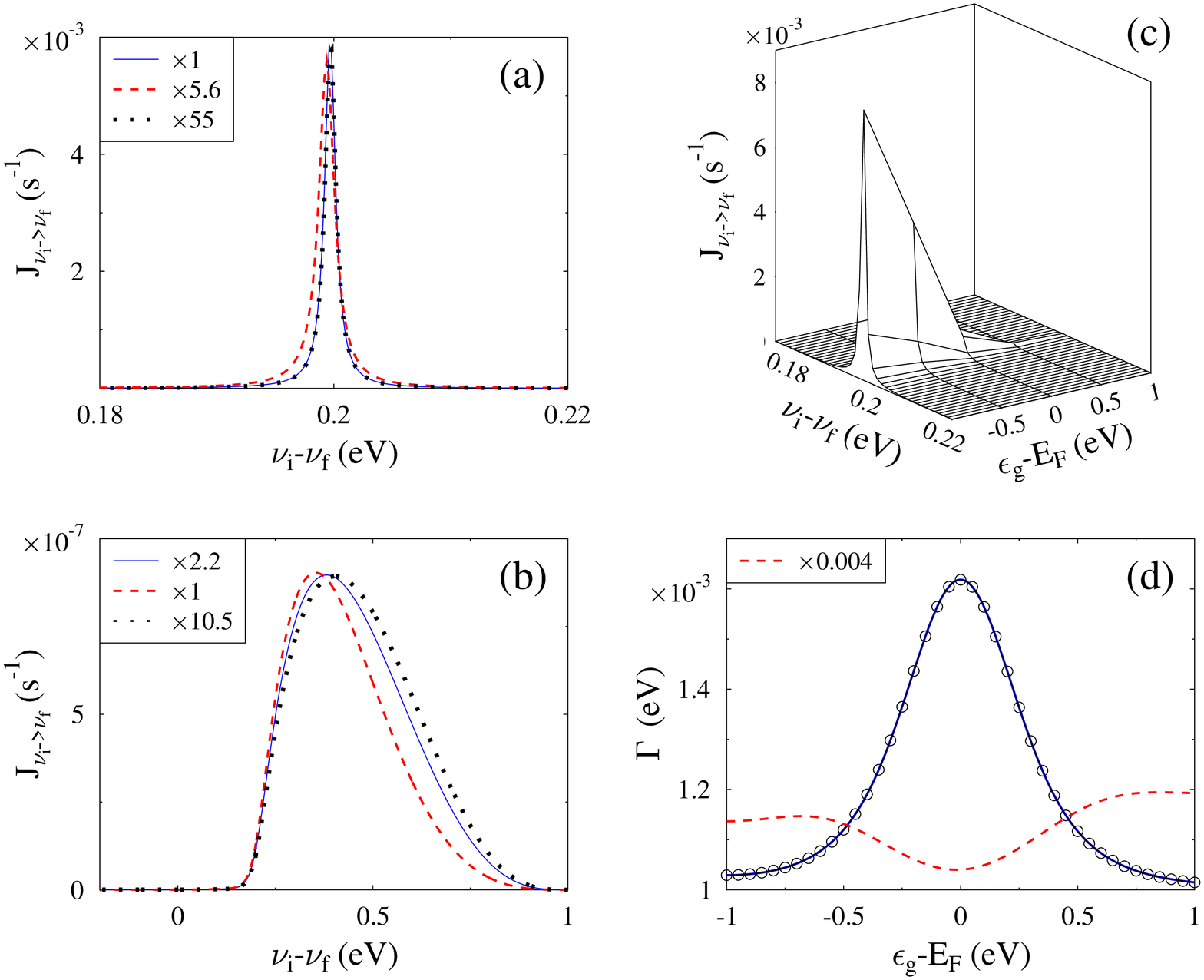}
\caption{\label{fig3}
(Color online) The Stokes component of the vibrational Raman scattering at equilibrium, 
$\mu_L=\mu_R=E_F$, for $\Gamma_m^K=0.4$~eV.
Shown are the contributions of (a)  Eq.~(\ref{Jnuifa}) and (b) Eq.~(\ref{Jnuifb}) vs. 
Raman shift  for three level positions: ($\varepsilon_g-E_F=-0.5$~eV, solid line, blue), 
at ($\varepsilon_g-E_F=0$, dashed line, red)
and above ($\varepsilon_g-E_F=0.5$~eV, dotted line, black).
The total Stokes signal, Eq.~(\ref{Jif}), as function of the Raman shift 
and level position is shown in panel (c). Panel (d) shows 
widths $\Delta\nu$ (standard deviations) of the two main contributions 
(Eq.~(\ref{Jnuifa}) - solid line, blue; Eq.~(\ref{Jnuifb}) - dashed line, red)  
as functions of the level position.
Circles indicate broadening of the molecular vibration due to coupling to electron-hole
excitations. 
See text for other parameters.
}
\end{figure}
%%%%%%%%%%%%%%%%%%%%%%%%%%%%%%%%%%%%%%%%%%%%%%%%%%%%%%%%%%%%%%%%%%%%%%%%%%%%%%%

Next, some simplification can be made by invoking the reasonable assumption that the
molecule-contacts coupling is much larger than the molecule-radiation field coupling 
as well as the electron-phonon interaction. 
Under this assumption we can disregard the latter interactions
in the expressions for the electronic Green functions,
taking the forms that correspond to a molecule coupled to the two metal leads
\begin{align}
 G_m^r(E) =& \bigg[E-\varepsilon_m+i\Gamma_m/2\bigg]^{-1}
 \\
 G_m^{<}(E) =& i\frac{\Gamma_m^L f_L(E)+\Gamma_m^R f_R(E)}
 {(E-\varepsilon_m)^2+(\Gamma_m/2)^2}
 \\
 G_m^{>}(E) =& -i\frac{\Gamma_m^L [1-f_L(E)]+\Gamma_m^R [1-f_R(E)]}
 {(E-\varepsilon_m)^2+(\Gamma_m/2)^2}
\end{align}
Here $\Gamma_m^K\equiv 2\pi\sum_{k\in K}\left\lvert V_{mk}\right\rvert^2\delta(E-\varepsilon_k)$ ($m=g,x$, $K=L,R$) is electron escape rate from level $m$ into 
contact $K$, $\Gamma_m=\Gamma_m^L+\Gamma_m^R$, 
$f_K(E)$ is the Fermi-Dirac thermal distribution in contact $K=L,R$.
 
  %%%%%%%%%%%%%%%%%%%%%%%%%%%%%%%%%%%%%%%%%%%%%%%%%%%%%%%%%%%%%%%%%%%%%%%%%%%%%%%
\begin{figure}
\centering\includegraphics[width=\linewidth]{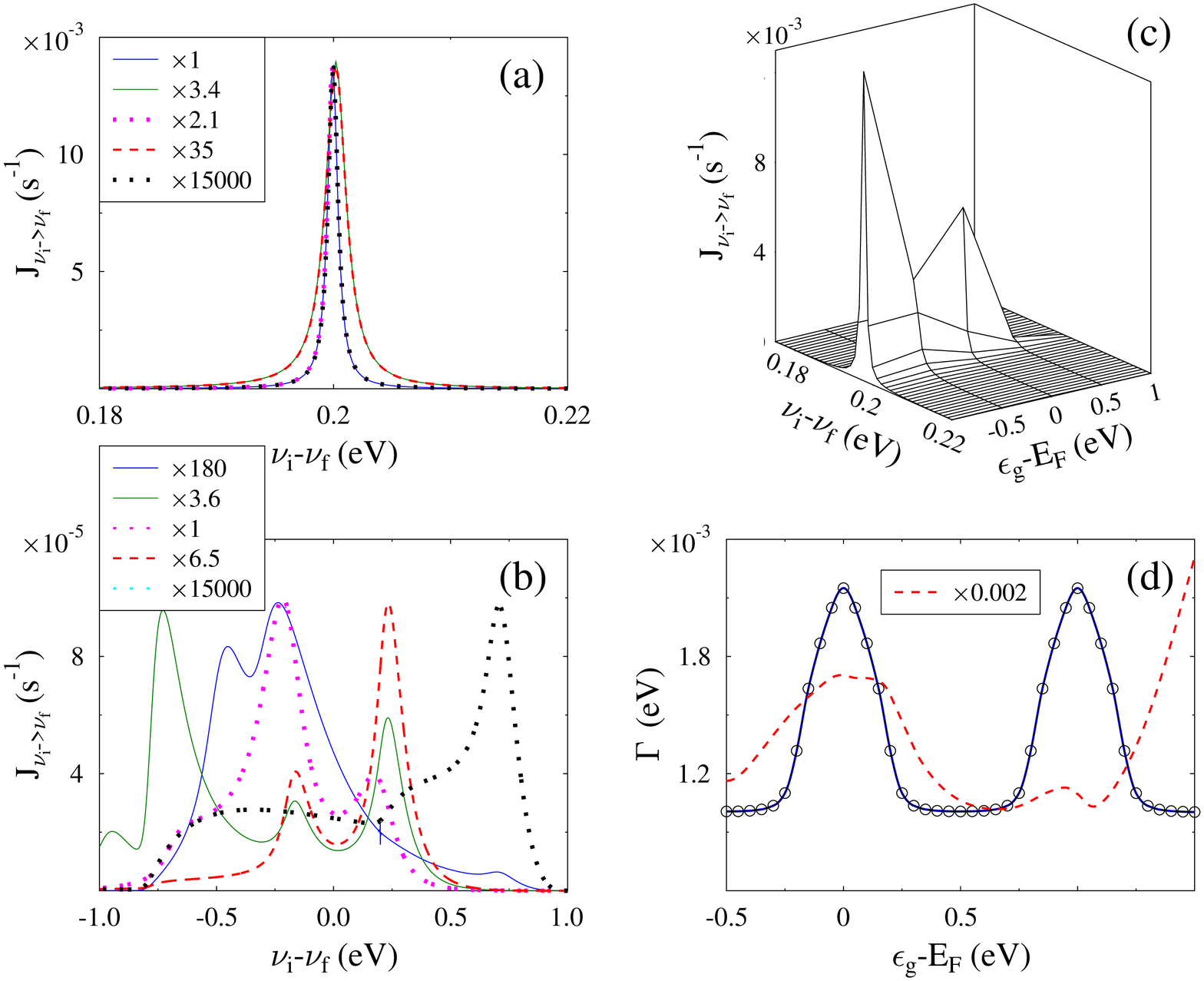}
\caption{\label{fig4}
(Color online) Stokes scattering from a biased junction, 
$\mu_L=0.5$~eV and $\mu_R=-0.5$~eV for $\Gamma_m^K=0.05$~eV.
Shown is the dependence of the Stokes signal on Raman shift for level position
$\varepsilon_g<\mu_R$ ($\varepsilon_g=-1$~eV, solid line, blue), 
$\varepsilon_g=\mu_R$ (solid line, green),
$\mu_R<\varepsilon_g<\mu_L$ ($\varepsilon_g=0$, dotted line, magenta),
$\varepsilon_g=\mu_L$ (dashed line, red), and
$\varepsilon_g>\mu_L$ ($\varepsilon_g=1$~eV, dotted line, black)
for the contributions of (a) Eq.~(\ref{Jnuifa}) and 
(b) Eq.~(\ref{Jnuifb}). The Inset shows the scaling parameters
used for the different lines.
The total Stokes scattering, Eq.~(\ref{Jif}), as function of the Raman shift 
 and level position is shown in panel (c). Panel (d) shows 
widths $\Delta\nu$ (standard deviations) of the two main contributions 
(Eq.~(\ref{Jnuifa}) - solid line, blue; Eq.~(\ref{Jnuifb}) - dashed line, red)  
as functions of the level position.
Circles indicate broadening of the molecular vibration due to coupling to electron-hole
excitations. 
See text for other parameters.
}
\end{figure}
%%%%%%%%%%%%%%%%%%%%%%%%%%%%%%%%%%%%%%%%%%%%%%%%%%%%%%%%%%%%%%%%%%%%%%%%%%%%%%%
  %%%%%%%%%%%%%%%%%%%%%%%%%%%%%%%%%%%%%%%%%%%%%%%%%%%%%%%%%%%%%%%%%%%%%%%%%%%%%%%
\begin{figure}
\centering\includegraphics[width=\linewidth]{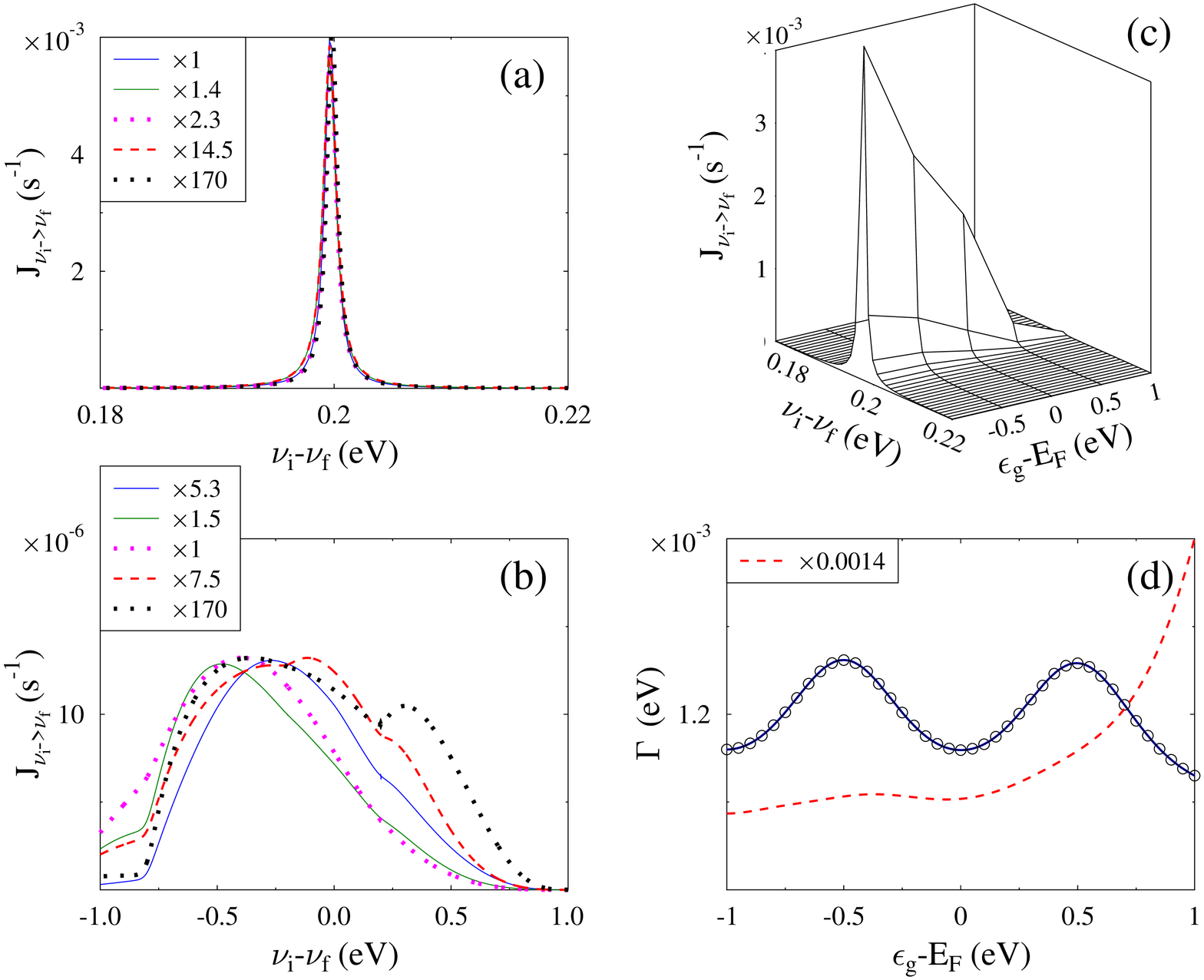}
\caption{\label{fig5}
(Color online) Stokes scattering from a biased junction, 
$\mu_L=0.5$~eV and $\mu_R=-0.5$~eV, for $\Gamma_m^K=0.4$~eV.
Shown is the dependence of the Stokes signal on Raman shift for level position
$\varepsilon_g<\mu_R$ ($\varepsilon_g=-1$~eV, solid line, blue), 
$\varepsilon_g=\mu_R$ (solid line, green),
$\mu_R<\varepsilon_g<\mu_L$ ($\varepsilon_g=0$, dotted line, magenta),
$\varepsilon_g=\mu_L$ (dashed line, red), and
$\varepsilon_g>\mu_L$ ($\varepsilon_g=1$~eV, dotted line, black)
for the contributions of (a) Eq.~(\ref{Jnuifa}) and 
(b) Eq.~(\ref{Jnuifb}). The Inset shows the scaling parameters
used for the different lines.
The total Stokes scattering, Eq.~(\ref{Jif}), as function of the Raman shift 
 and level position is shown in panel (c). Panel (d) shows 
widths $\Delta\nu$ (standard deviations) of the two main contributions 
(Eq.~(\ref{Jnuifa}) - solid line, blue; Eq.~(\ref{Jnuifb}) - dashed line, red)  
as functions of the level position.
Circles indicate broadening of the molecular vibration due to coupling to electron-hole
excitations. 
See text for other parameters.
}
\end{figure}
%%%%%%%%%%%%%%%%%%%%%%%%%%%%%%%%%%%%%%%%%%%%%%%%%%%%%%%%%%%%%%%%%%%%%%%%%%%%%%%

For the evaluation of the phonon Green functions we again disregard 
the molecule-radiation field coupling, but keep the electron-phonon interaction.
This leads to 
\begin{align}
\label{Dr}
D^r(\omega) &= \bigg[\left[D_0^r(\omega)\right]^{-1}-\Pi_{th}^r(\omega)-\Pi_{el}^r(\omega)\bigg]^{-1}
\\
\label{Dgtlt}
D^{>/<}(\omega) &= 
D^r(\omega) \bigg(\Pi_{th}^{>/<}(\omega)+\Pi_{el}^{>/<}(\omega)\bigg) D^a(\omega)
\end{align}
We will henceforth assume that $\omega>0$ and use $D^{>/<}(-\omega)=D^{</>}(\omega)$ 
to access the $\omega<0$ region.
In Eqs.~(\ref{Dr}) and (\ref{Dgtlt}) $D^a(\omega)=[D^r(\omega)]^{*}$, 
\begin{equation}
 D_0^r(\omega) = \frac{1}{\omega-\omega_v+i\delta}
 -\frac{1}{\omega+\omega_v+i\delta}
\end{equation} 
is the retarded projection of free phonon Green function, and
\begin{align}
\Pi_{th}^r(\omega) =& -i\frac{\gamma(\omega)}{2}
\\
\Pi_{th}^{<}(\omega) =& -i\gamma(\omega)N(\omega)
\\
\Pi_{th}^{>}(\omega) =& -i\gamma(\omega)[1+N(\omega)] 
\end{align}
are the projections of the self-energy of the molecular vibration due to its coupling 
to the (bosonic) white % ????
thermal bath.
Here $N(\omega)$ is the Bose-Einstein thermal distribution and
$\gamma(\omega)=2\pi\sum_\beta \left\lvert V^{th}_\beta\right\rvert^2\delta(\omega-\omega_\beta)$ is the dissipation rate of molecular vibrational excitation 
due to coupling to thermal bath.
The self energy of the molecular phonon associated with the electron-vibration coupling
is treated at the level of the Born approximation
\begin{align}
 \Pi_{el}^r(\omega) =& -i M_g^2\int\frac{dE}{2\pi}\,
 \bigg(G_g^{<}(E)\, G_g^a(E-\omega)
 \\ &\qquad\qquad\qquad
 +G_g^r(E)\, G_g^{<}(E-\omega)\bigg)
 \nonumber \\
 \label{Piellt}
 \Pi_{el}^{<}(\omega) =& -i M_g^2 \int\frac{dE}{2\pi}\, G_g^{<}(E)\, G_g^{>}(E-\omega)
 \\
 \label{Pielgt}
 \Pi_{el}^{>}(\omega) =& -i M_g^2 \int\frac{dE}{2\pi}\, G_g^{>}(E)\, G_g^{<}(E-\omega)
\end{align}
Before describing our numerical results, it is important to note the different physical 
origins of the four contributions, Eqs.~(\ref{Jnuifa})-(\ref{Jnuifd}), to the Raman signal, 
that can be inferred from the different forms of the electronic Green functions appearing 
in them and the forms of the corresponding energy denominators. 
It is convenient to look at them in comparison to the pure electronic Raman components
discussed in Ref.~\cite{MGANPRB11} (see Fig. 2 in this reference). 
Without the vibrational shift the contribution (\ref{Jnuifa})  with
$G_g^{<}\, G_g^{<}\, G_x^{>}\, G_x^{>}$ would be the Rayleigh line where 
each scattering event involves a single electron-hole pair
- an occupied electronic level near $E_g$ and an empty electronic level near $E_x$. 
The contribution (\ref{Jnuifb}) with $G_g^{<}\, G_g^{>}\, G_x^{>}\, G_x^{>}$
if considered without the vibrational shift corresponds to that contribution to the pure
electronic Raman scattering where the difference between the initial and final photon 
energy is  expressed by moving an electron between two metal levels close to $E_g$, 
requiring one of these levels to be occupied and the other empty. 
The term (\ref{Jnuifc}) that depends on $G_g^{<}\, G_g^{<}\, G_x^{<}\, G_x^{>}$
is similar, except that the difference between incoming and outgoing photons is 
expressed in electron motion between two levels near $E_x$, 
again requiring one of them to be occupied and the other empty. 
Finally, the contribution (\ref{Jnuifd}) that contains $G_g^{<}\, G_g^{>}\, G_x^{<}\, G_x^{>}$
splits the photon energy difference between two electronic transitions, one
near $E_x$ and the other near $E_g$.

Two observations follow, still on this qualitative level: 
First, in equilibrium and at low bias, in the common situation where the lower and 
upper electronic orbitals are far below and far above the metal(s) Fermi energy(ies) respectively, Eq.~(\ref{Jnuifa}) will be the dominant contribution to the
vibrational Raman signal. 
Second, the vibrational Raman lines associated with this contribution will be
narrow in the sense that their width will not reflect the excitation of electron-hole pairs 
in the metal. The contributions (\ref{Jnuifb}) and (\ref{Jnuifc}) will be important 
in situations where, respectively, $E_g$ and $E_x$ are close to
the metals Fermi energies. Furthermore, these contributions will be considerable broader, 
reflecting the excitation of electron-hole pairs in the metal alongside 
the vibrational excitation. Note that at low temperatures this broadening will be asymmetric, 
corresponding to an electronic side-band of the vibrational Raman transition as recently discussed in Ref.~\cite{ApkarianMGANPRB16}. Finally, we expect
that also the pure vibrational Raman spectrum associated with 
Eq.~(\ref{Jnuifa}) will be broader when one of the the molecular electronic levels 
is close to the Fermi energy, because of the increased importance of
the electronic relaxation channel for the molecular vibration in this 
situation.\cite{KaasbjergNitzanPRB13}

%%%%%%%%%%%%%%%%%%%%%%%%%%%%%%%%%%%%%%%%%%%
 
\section{Numerical results}\label{numres}
Here we present numerical results for the Raman flux, Eq.~(\ref{Jnu}), 
for the model (\ref{H})-(\ref{V}). Below we focus on the most prominent contributions,
Eqs.~(\ref{Jnuifa}) and (\ref{Jnuifb}).
The following parameters are used in these calculations:
$T=100$~K, $\varepsilon_x-\varepsilon_g=2$~eV 
(the absolute level positions are varied as described below), 
$\Gamma^L_m=\Gamma^R_m=0.05$~eV (in Figs.~\ref{fig2} and \ref{fig4})
and $0.4$~eV (in Figs.~\ref{fig3} and  \ref{fig5}), ($m=g,x$),
$\omega_v=0.2$~eV, $\gamma(\omega_v)=10^{-3}$~eV, and $M_g=0.03$~eV.
The Fermi energy is chosen as the origin, $E_F=0$, and the bias is applied symmetrically
$\mu_L=E_F+|e|V_{sd}/2$ and $\mu_R=E_F-|e|V_{sd}/2$.
The incident frequency is taken as $\nu_i=1$~eV, which corresponds
for the present choice of molecular parameters to off-resonant Raman scattering. 
The couplings to the radiation field are assumed to satisfy
$U_{\alpha D}^{(0)}=U_{\alpha D}^{(1)}=0.01$~eV.
The optical resolution windows of the incident energy and measuring device, 
$\Delta\nu_i$ and $\Delta\nu_f$ in Eq.~(\ref{Jnu}), 
are taken to be the same, $0.01 eV$.
The calculations were performed on an energy grid
spanning the range from $-5$ to $5$~eV with step size $5\times 10^{-5}$~eV.

We envision an experiment in which the position of the molecular resonances 
can be changed by a gate voltage. We start from the situation where level $g$ 
is far below the Fermi energy and level $x$ is far above it,
so that the lower level is occupied and upper one is empty, and
consider the effect on the Raman spectrum of applying a gate voltage 
to move $\varepsilon_g$ to the vicinity of, and then beyond, the chemical potentials.
In this regime the two main contributions to the Raman flux are given by 
Eqs.~(\ref{Jnuifa}) and (\ref{Jnuifb}) with the first one dominating the intensity of the 
Stokes line. (As explained above, the terms (\ref{Jnuifc}) and (\ref{Jnuifd}) 
are potentially important only when the excited state is populated).
The Raman linewidths reported below are estimated using the standard deviation 
associated with the corresponding Raman peak calculated on the employed energy grid.

Figure~\ref{fig2} shows results of of this calculation for the equilibrium case, 
$\mu_L=\mu_R=E_F$.
Note that the intensity of the Stokes line decreases with decrease of the population in 
the ground state (see Fig.~\ref{fig2}c),
however the implication of this observation should be understood with respect to 
the 2-level model used here. In reality, when $\varepsilon_g$ goes up and above 
the metal Fermi energy, other lower molecular levels will contribute to the Raman signal.
Disregarding this issue, the following additional observations can be made:
\begin{enumerate}
\item The dominant Raman feature is indeed that associated with contribution 
(\ref{Jnuifa}) Ð the electronically elastic/vibrationally inelastic signal. 
The contribution (\ref{Jnuifb}) becomes comparable when $\varepsilon_g$ is near 
the metal Fermi energy. It should be kept in mind that and additional broad feature, 
the electronically inelastic/vibrationally elastic (pure electronic) is not displayed 
in these figures. In experimental spectra, the signal (\ref{Jnuifb}) may often become 
part of this broad electronic background.
\item The width of the contribution (\ref{Jnuifb}) is far greater than that of 
the electronically elastic term (\ref{Jnuifa}), 
as long as $\varepsilon_g$ is far from the metal Fermi energy. 
However, the width of (\ref{Jnuifa}) increases considerably
when $\varepsilon_g$ approaches $E_F$.
\item The widths of the two contributions, (\ref{Jnuifa}) and (\ref{Jnuifb}),  
behave symmetrically about the Fermi energy (see Figs.~\ref{fig2}a and b).
Such behavior is expected since in both cases the width is defined by  convolution
of electron and hole populations, $G_g^{<}(E_{g1})\, G_g^{>}(E_{g2})$, 
which at equilibrium is symmetric relative to  the Fermi energy. 
\item Comparing the results displayed in Figures~\ref{fig2} and \ref{fig3} 
(small and large molecule-metal coupling ($\Gamma$), respectively, 
we note that the dominant low bias feature, namely the contribution 
(\ref{Jnuifa}) is essentially the same in both cases. 
Interestingly, when $\varepsilon_g$ is at the Fermi energy 
(dashed red lines in Figs.~\ref{fig2}a and \ref{fig3}a), this feature is broader in 
the smaller $\Gamma$ case. This is also seen in comparing 
Figs.~\ref{fig2}d and \ref{fig3}d. This behavior reflects the fact that 
when $\Gamma>k_BT$, even when $\varepsilon_g=E_F$,
most of the electronic spectral density (of width $\Gamma$) is outside 
the region of partial electronic occupation ($f(1-f)\neq 0$ where f is the Fermi distribution) 
in which the electronic channel for vibrational relaxation is open. 
The fact that the spectra in Fig.~\ref{fig3} are smoother and less structured than 
in Fig.~\ref{fig2} similarly reflects the fact that for large $\Gamma$ all behaviors 
associated with the position of $\varepsilon_g$ relative to $E_F$ and the width of 
the partially populated region are smoothened.
\end{enumerate}

The width of the vibrational Raman lines reflects three types of contributions. 
First there is the relaxation to the thermal bosonic environment that is not affected 
(in our model) by the bias and gate potentials. Second is the additional relaxation 
channel due to electron-vibration coupling, that can dominate the overall width 
when the molecular electronic level approaches the metal Fermi level.  
The structure of this contribution suggests that the width of the term 
(\ref{Jnuifa})  (solid line in Fig.~\ref{fig2}d) is dominated by 
the (renormalized) density of molecular vibration (circles in Fig.~\ref{fig2}d). 
Finally, as discussed above, there is the electron-hole sideband
that appears prominently in the term (\ref{Jnuifb}) 
(as well as (\ref{Jnuifc}) and (\ref{Jnuifd})). Note again that in actual observations it will
not be easy to distinguish between this sideband to the vibrational transition 
and the underlying Raman continuum that originates primarily from the pure 
electronic Raman scattering.\cite{MGANPRB11}

We now turn to the nonequilibrium situation with $\mu_L=0.5$~eV and $\mu_R=-0.5$~eV.
The total Stokes intensity is here affected by two factors:
the population of the lower level and the current induced heating of the 
molecular vibration. As a result, the decrease in the Stokes intensity
when $\varepsilon_g$ approaches the lowest chemical potential  
due to depletion of the level population 
changes to increase in the intensity when the level is in the bias window 
(nonequilibrium feature) - see Fig.~\ref{fig4}c. 
The width of the dominant contribution (\ref{Jnuifa}) shows similar behavior as in 
the equilibrium case, with increase of the width resulting from opening the
electronic relaxation channel when $\varepsilon_g$ approaches the metal Fermi 
energies. This contribution to the width is again symmetric about each of 
the Fermi energies (see Figs.~\ref{fig4}a and \ref{fig3}d).
In contrast, the nonequilibrium electronic distribution, 
in particular the existence of two energy regions of partial populations of 
metal electronic states, causes drastic changes and more structure in both 
lineshape (Fig.~\ref{fig4}b) and linewidth (Fig.~\ref{fig3}d) of the contribution
(\ref{Jnuifb}) as compared to equilibrium case. 
This structure is again smoothened in the large $\Gamma$ case (Fig.~\ref{fig5}).
Still, since this peak is much lower and broader than that of (\ref{Jnuifa}),
it may be considered as part of the electronic Raman background.

\section{Conclusion}\label{conclude}
Within a simple two-level model of a molecular junction we 
consider off-resonant Raman scattering and discuss 
dependence of Stokes linewidth on gate and bias voltages.
We focus on low bias regime, where upper level is almost empty, and
thus consider only `normal Raman' contribution to the total signal
(i.e. Raman scattering which originates at the lower molecular level). 
Employing realistic parameters we show that the linewidth changes non-monotonically
with gate voltage demonstrating maximum at resonance between
molecular level and chemical potential(s) of metallic contacts. 
Analysis shows that the effect is due to opening of an electronic relaxation channel
for molecular vibrations by which e-h excitations are formed in metallic contacts.
At low biases and for realistic parameters this mechanism is the dominant contribution 
to the Stokes linewidth. Other mechanisms are relaxation of molecular vibration due to
coupling to the thermal environment  and surface plasmons. The latter was not included
in the consideration due to mismatch between characteristic plasmons 
and molecular vibrations frequencies. Note that the model also disregards
inhomogeneous broadening and pure dephasing contributions. 
Experimental verification of our theoretical prediction seems to be 
a realistic possibility.

%%%%%%%%%%%%%%%%%%%%%%%%%%%%%%%%%%%%%%%%%%%%%%%%%%%%%%%%%%%%%%%%%%%%%%%%%%%%%%%
\begin{acknowledgments}
The research of AN is partially supported by the Israel Science Foundation 
and  the US-Israel Binational Science Foundation.
MG gratefully acknowledges support by the US Department of Energy.
\end{acknowledgments}
%%%%%%%%%%%%%%%%%%%%%%%%%%%%%%%%%%%%%%%%%%%%%%%%%%%%%%%%%%%%%%%%%%%%%%%%%%%%%%%
%\bibliography{rlwidth}

%merlin.mbs apsrev4-1.bst 2010-07-25 4.21a (PWD, AO, DPC) hacked
%Control: key (0)
%Control: author (8) initials jnrlst
%Control: editor formatted (1) identically to author
%Control: production of article title (-1) disabled
%Control: page (0) single
%Control: year (1) truncated
%Control: production of eprint (0) enabled
%

\end{document}